# Defect Induced Room Temperature Ferromagnetism in High Quality Co-doped ZnO Bulk Samples

M. P. F. de Godoy,[a] X. Gratens,[b] V. A. Chitta,[b] A. Mesquita,[c] M. M de Lima Jr.,[d] A. Cantarero,[e] G. Rahman,[f†] J. M. Morbec,[g] and H. B. de Carvalho[h]*

[a] *Departamento de Física, Universidade Federal de São Carlos - UFSCar, 13565-905 São Carlos, Brazil.*

[b] *Instituto de Física, Universidade de São Paulo - USP, 05508-090 São Paulo, Brazil.*

[c] *Departamento de Física, Instituto de Geociências e Ciências Exatas, Universidade Estadual Paulista – UNESP, 13500-900 Rio Claro, Brazil.*

[d] *Materials Science Institute, University of Valencia, 22085, E-46071 Valencia, Spain.*

[e] *Institute of Molecular Science, University of Valencia, 22085, E-46071 Valencia, Spain.*

[f] *Department of Physics, Quaid-i-Azam University, 45320 Islamabad, Pakistan.*

[g] *School of Chemical and Physical Sciences, Keele University, Keele ST5 5BG, United Kingdom*

[h] *Universidade Federal de Alfenas - UNIFAL, 37130-000 Alfenas, Brazil.*

*Corresponding Authors*:
† gulrahman@qau.edu.pk.
* hugo.carvalho@unifal-mg.edu.br.

**ABSTRACT:** The nature of the often reported room temperature ferromagnetism in transition metal doped oxides is still a matter of huge debate. Herein we report on room temperature ferromagnetism in high quality Co-doped ZnO ($Zn_{1-x}Co_xO$) bulk samples synthesized via standard solid-state reaction route. Reference paramagnetic Co-doped ZnO samples with low level of structural defects are subjected to heat treatments in a reductive atmosphere in order to introduce defects in the samples in a controlled way. A detailed structural analysis is carried out in order to characterize the induced defects and their concentration. The magnetometry revealed the coexistence of a paramagnetic and a ferromagnetic phase at room temperature in straight correlation with the structural properties. The saturation magnetization is found to increase with the intensification of the heat treatment, and, therefore, with the increase of the density of induced defects. The magnetic behavior is fully explained in terms of the bound magnetic polaron model. Based on the experimental findings, supported by theoretical calculations, we attribute the origin of the observed defect-induced-ferromagnetism to the ferromagnetic coupling between the Co ions mediated by magnetic polarons due to zinc interstitial defects.

# 1 INTRODUCTION

Spintronics, the spin-based electronics, is a new class of technological devices that take advantage of the spin degree of freedom of carriers. The most significant examples of successful commercial spintronic devices are the magnetic sensors (spin valves), and the non-volatile magnetic memories (magnetic tunnel junctions). It is worth stressing that both of these devices are based on magnetic metallic alloys. Besides, it was demonstrated that semiconductor spintronic devices can offer new functionalities, such as the electric control of the magnetization and of the Curie temperature [1]. These functionalities cannot be implemented in the metal-based spintronic or, even worse, with the conventional electronics devices. Unfortunately, semiconductor spintronic devices are still commercially unavailable, despite of the last decades efforts [2]. The main problem to be overcome is the lack of a room temperature ferromagnetic semiconductor that could be used in a practical device, such as a spin field emission transistor (spin-FET) [3, 4]. The most promising candidates to solve this problem are semiconductors that, when doped with magnetic impurity atoms, display ferromagnetism, the so called dilute magnetic semiconductors (DMSs).

In 2000 Dietl et al. [5] theoretically predicted that Mn-doped ZnO and GaN (large band gap semiconductors) would present room-temperature ferromagnetism (RTFM), which sparked a major worldwide rush to develop a truly DMS, with special attention devoted to dilute magnetic oxides (DMOs), more specifically to transition-metal (TM) doped ZnO [6]. The experimental and theoretical results earlier obtained on the magnetic properties of the DMOs are highly controversial. TM-doped ZnO has been found to present RTFM [7], but it was also true for ZnO doped with non-magnetic elements, like carbon and Li [8, 9], and even for undoped ZnO with specific structural defects [10, 11]. However, it is clear now that the structural defects play an important role to tune the functional properties of the oxides, particularly their magnetic properties. The mentioned controversy can be mainly attributed to the lack of knowledge of the nature and densities of defects present in the studied samples. We believe that special attention has not been paid to control specific parameters during the preparation of the materials and, therefore, a concise effort is necessary to address this issue.

In this context, we report here a detailed study on the correlation between the structural and magnetic properties of Co-doped ZnO ($Zn_{1-x}Co_xO$) bulk samples. High quality samples with low level of structural defects (with high crystallinity) were prepared following the procedures published before [12]. The samples were then subjected to heat treatments in reductive atmosphere in order to introduce specific structural defects in a controlled manner. Special attention was also paid to detect the presence of alternative sources of ferromagnetism, such as secondary phases and nanocrystals embedded in the bulk material. First-principles calculations based on density functional theory (DFT) were also performed to support the experimental results. We demonstrate that the induced ferromagnetism can be entirely explained in terms of the bound magnetic polaron (BMP) model [13] associated with a zinc interstitial ($Zn_i$) shallow-donor defect band.

## 2 EXPERIMENTAL AND THEORETICAL METHODS

Polycrystalline $Zn_{0.96}Co_{0.04}O$ bulk samples were prepared by the standard solid-state reaction method. Stoichiometric amounts of ZnO (99.998%) and $Co_3O_4$ (99.7%) were mixed and ball milled for 5 h using Zn spheres. The resulting powder was cold compacted by an uniaxial pressure of 600 MPa in the form of pellets (green pellets). The green pellets were finally sintered in oxygen atmosphere for 4 h at 1400 °C (labelled as *AP* sample). Thereafter, one batch of the oxygen sintered samples was heat treated in a gaseous mixture of argon (95%) and hydrogen (5%) for 3 hours at 600 °C (labelled as *H*600) and another batch was also heat treated for 3 hours but at 800 °C (labelled as *H*800). We also present the result for an undoped ZnO sample prepared at the same conditions of sample *AP*. The effects of heat treatment on the structural properties were investigated via X-ray diffraction (XRD) using a Philips X-ray diffractometer employing $FeK_\alpha$ radiation and a graphite monochromator. XRD data were recorded in the range of $2\theta = 35°-65°$ with steps of 0.01° at 1.2 s/step. Structural analysis was performed using the Rietveld method as implemented in the General Structure Analysis System (GSAS) software package with the graphical user interface EXPGUI [14, 15] (Ref. ICSD: 97-957). The structure and the elemental spatial distribution were evaluated via high-resolution transmission electron microscopy (HRTEM), selected area electron diffraction (SAED) and energy dispersive X-ray spectrometry (EDS). These analyses were conducted on cross-sectional samples prepared by standard mechanical polishing followed by $Ar^+$-ion milling (Fishione 1010) by using a Field Emission Gun (FEG) TECNAI G2 F20 microscope operated at 200 kV. Raman spectroscopy was used to study the Co incorporation on the ZnO matrix and the resulting lattice disorder, as well as to analyze the formation of segregated secondary phases. Raman measurements were carried out at room temperature using a Jobin-Yvon-64000 micro-Raman system in the backscattering geometry and the 514, 488 and 466 nm lines of an Ar-Kr laser for excitation. Resonant Raman scattering analysis was performed using a He-Cd excitation laser with a 325 nm (3.815 eV) line, which is around 440 meV above the wurtzite ZnO (*w*-ZnO) energy gap. Co *K*-edge X-ray absorption near-edge structure (XANES) and extended X-ray absorption fine structure (EXAFS) were used to determine the valence state and to evaluate the environment of Co in the ZnO lattice. Normalized XANES spectra were extracted with the MAX-Cherokee code. The sample thicknesses were optimized by the Multi-Platform Applications for XAFS (MAX) software package Absorbix code [16]. The theoretical XANES spectra were calculated by the FEFF9 ab initio code [17, 18] whose input files were issued from MAX-Crystalffrev software, which takes into account substitution disorder and random vacancies in the structure [16]. EXAFS spectra were extracted with the MAX-Cherokee code while the fitting procedure and comparison between experimental and theoretical EXAFS curves were conducted with the MAX-Roundmidnight package. The theoretical EXAFS spectra were calculated by the FEFF9 ab initio code [17, 18] whose input files were issued from MAX-Crystalffrev software. In our work the relevant measure of the fit quality, the reduced statistical $\chi^2$, is named *QF* (quality factor). X-ray absorption measurements were taken in the transmittance mode at the XAFS2 beamline from the Brazilian Synchrotron Light Laboratory (LNLS),

Campinas, Brazil. Changes in the density of defects were estimated by Photoluminescence (PL) and *dc* electrical measurements. PL spectra were recorded at 5 K and excited using a He-Cd laser (325 nm). The electrical properties of the prepared samples were evaluated by measuring the conductivity and Hall effect with the samples contacted in the van der Pauw geometry. Magnetic characterization was carried out using a standard superconducting quantum interference device magnetometer (SQUID).

The structural, electronic, and magnetic properties of the Co-doped samples were also investigated by means of *first-principles* calculations based on density functional theory [19]. Spin-polarized calculations were performed with local density approximation (LDA) [20, 21], norm-conserving Troullier-Martins pseudopotentials [22], using the SIESTA code [23]. We used an energy cutoff of 200 Ry and a double-zeta basis set with polarization function for all atoms. To study the electronic origin of magnetism of Co-doped ZnO, we used a 3 × 3 × 3 supercell (108 atoms) and considered intrinsic defects in ZnO, i.e., Zn interstitial ($Zn_i$) and Zn vacancy ($V_{Zn}$), as well as Co impurity at Zn site. We also considered the interaction of Co with these intrinsic defects following our previous procedure [24].

## 3  RESULTS AND DISCUSSION

### 3.1 *X-ray Diffraction*

Figure 1 shows the experimental XRD patterns and the theoretical Rietveld plot for the whole set of samples. The difference between the experimental and theoretical data is also shown. All the observed peak positions match the powder diffraction data for polycrystalline *w*-ZnO, space group *P6₃mc*, with no indication, within the XRD detection limit, of the presence of additional secondary Co phases. The narrow linewidths of the diffraction peaks reveal the good crystallinity quality of the samples. Rietveld analysis was conducted by placing the $Zn^{2+}$ and $O^{2-}$ ions located at the initial positions (1/3, 2/3, 0) and (1/3, 2/3, z), respectively. It is possible to observe the good agreement between the observed and calculated pattern. All the obtained lattice parameters are listed in Table 1. These data are quite similar to those reported for undoped $w$-ZnO [25], and there is no indication of lattice distortions caused by the Co incorporation and the heat treatments over the samples. This would be expected, since four-coordinated $Co^{2+}$ ions present ionic radius of 0.58 Å, a value quite close to that of the $Zn^{2+}$ ions in the $w$-ZnO lattice (0.60 Å) [26]. This result indicates that the Co assumes a tetrahedral coordination and a +2 oxidation state, which highlights the Zn substitutional character of the Co-doping at the $w$-ZnO matrix in the studied samples.

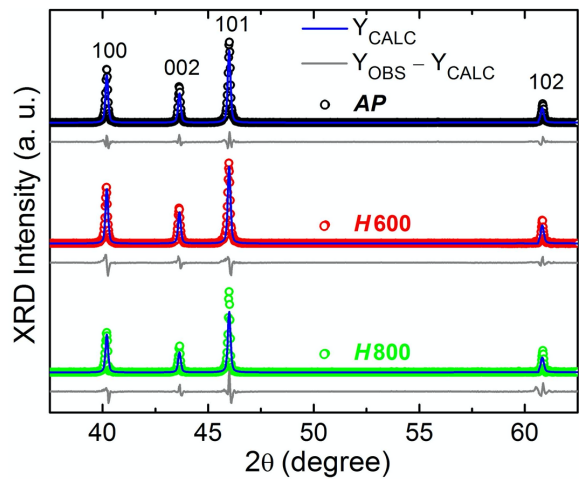

**Figure 1.** Refined XRD pattern for the polycrystalline $Zn_{0.96}Co_{0.04}O$ bulk samples. The observed and the calculated profiles are represented by open symbols and solid line curves, respectively. The lower curve is the difference plot. The corresponding refinement results are presented in Table 1.

**Table 1.** The refined structural parameters for the $Zn_{0.96}Co_{0.04}O$ samples. $\chi^2$ is the square of Goodness of fit indicator, and $R_{WP}$ is the refinement quality parameter.

| Sample | $a$ (Å) | $c$ (Å) | $c/a$ | Volume (Å³) | $\chi^2$ | $R_{WP}$ |
|---|---|---|---|---|---|---|
| AP | 3.24979(4) | 5.2049(9) | 1.6016 | 47.606(2) | 3.21 | 0.179 |
| H600 | 3.25131(6) | 5.2064(1) | 1.6013 | 47.664(2) | 4.05 | 0.177 |
| H800 | 3.24926(5) | 5.2027(1) | 1.6012 | 47.570(2) | 2.06 | 0.181 |

### 3.2 Electron Microscopy and Elemental Analysis

In the context of the DMOs, a careful analysis concerning the spatial elemental distribution of dopants over the oxide matrix is important in order to exclude the presence of secondary or segregated magnetic components (like metallic cobalt clusters in the present case) that can cause ferromagnetic signals and lead us to erroneously conclude for the observation of intrinsic ferromagnetism. Figures 2(a) and (b) show a representative TEM and HRTEM micrographs for the *H*800 sample, respectively. At this scale, no evidences of segregated secondary components were detected for all the samples. The SAED analysis (Fig. 2(c)) indicates a hexagonal *w*-ZnO structure (zone axis $1\bar{2}1\bar{3}$). No satellite diffraction spots were detected, confirming the single-crystal nature of the structure of the samples. Representative elemental maps of the sample area in Fig. 2(d) are presented in Fig. 2(e) and Fig. 2(f). The random distribution of Co ions is quite homogeneous in all samples as seen in Fig. 2(f) for sample *H*800 and, again, no indication of cobalt-rich regions was observed. The measured effective cobalt concentration ($x_E$) for the samples was $x_E = 0.0405(3)$.

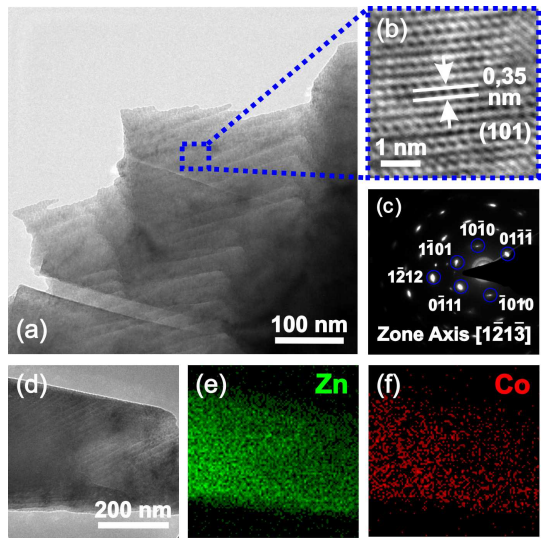

**Figure 2.** (a) Representative TEM and (b) HRTEM micrographs for the $Zn_{0.96}Co_{0.04}O$ bulk sample heat treated for 3 hours at 800 °C (*H*800). (c) SAED patterns of the enclosed area in (b). (d) TEM micrography and correspondent (e) zinc and (f) cobalt elemental mapping obtained by EDS.

### 3.3 *Raman Scattering Spectroscopy*

Raman spectra from our set of samples are shown in Fig. 3(a). A series of narrow peaks centered at 330, 380, 410 and 438 cm$^{-1}$ was observed. They are assigned to the $2E_{2L}$ at the M-point of the Brillouin zone, $A_1$(TO), $E_1$(TO) and $E_{2H}$ modes, respectively. [27, 28]. The spectra are normalized by the integrated intensity of the $E_{2H}$ mode. A significant result from the Raman data is the complete absence of peaks related to segregated secondary phases, as it has been observed for some Co-doped ZnO samples. The most common secondary phases are CoO and $Co_3O_4$ [29, 30]. For both materials the main Raman mode should appears at ~690 cm$^{-1}$ [31]. We do not observe, however, any indication of a peak at this frequency for our set of samples. Therefore, Raman scattering results corroborate the XRD and the TEM results, as they indicate the absence of segregated secondary phase in the studied samples. We also observed a broad band at ~500-600 cm$^{-1}$ that encloses several peaks, where the most prominent ones are centered at ~550 cm$^{-1}$ (addressed as S') and other at ~573 cm$^{-1}$ (addressed as LO). The peak at ~573 cm$^{-1}$ is attributed to the mixture of the $A_1$(LO) and $E_1$(LO) modes [27]. However, low temperature off-resonant Raman measurements (not shown) reveal that the LO mode can be mainly associated with the $A_1$(LO) mode in our samples. The nature of these two modes, S' and LO, can be revealed by approaching a resonant condition. The inset of Fig. 3(a) presents the spectra obtained for the *AP* sample under excitation wavelength of 514, 488 and 466 nm. As compared to the main mode $E_{2H}$, the intensities of both S' and LO modes are increased. This behavior is an indicative of resonant Raman scattering and its observation using sub band gap excitation is consistent with a model developed for extrinsic Frölich interactions [32, 33]. The Co doping and the structural defects due to heat treatment process introduce electronic levels within the band gap which would be required for the resonant Raman scattering observation. This assumption is further confirmed by the analysis of the photoluminescence and the *dc* electrical characterization results. Thus, we can conclude that the origin of the S' and LO modes is related to the doping and also to structural defects introduced via heat treatment. Table 2 presents the wavenumber (ω) and linewidth (λ) of the main mode $E_{2H}$ and the integrated intensity ratio of S' and LO to $E_{2H}$ obtained by fitting of the modes by the sum of Lorentzian shape functions. We observe that the heat treatment induces a slight redshift and a broadening of the $E_{2H}$, the intensities of S' and LO mode with respect to the $E_{2H}$ intensity also increase. These features reveal an increased lattice disorder associated with structural defects due to the heat treatment since all the samples have the same Co content.

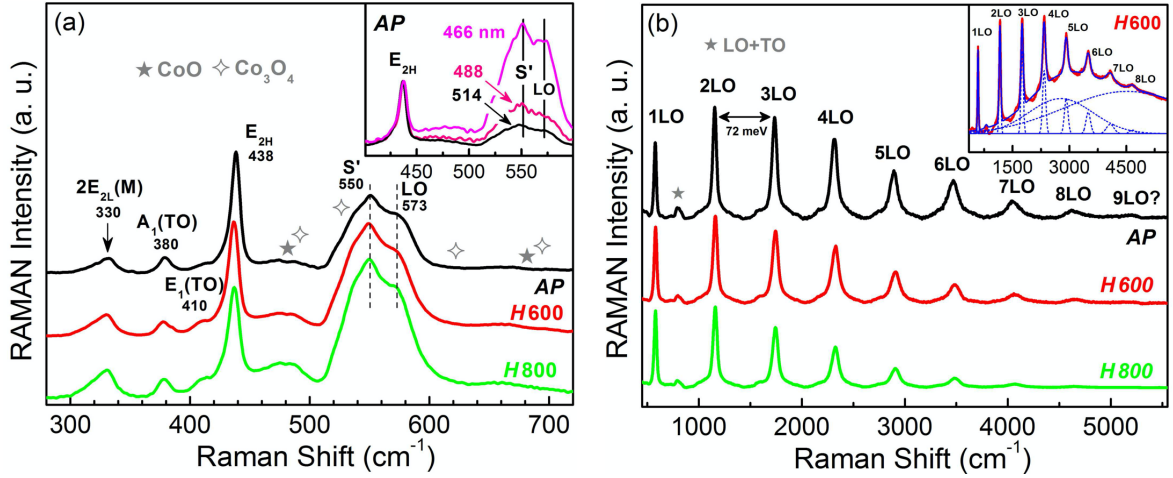

**Figure 3.** (a) Raman scattering spectra for the polycrystalline $Zn_{0.96}Co_{0.04}O$ bulk samples obtained at room temperature at excitation of 488 nm. The symbols mark the positions of potential secondary phases CoO and $Co_3O_4$. The inset shows the recorded spectra for the *AP* sample under excitation wavelengths of 514, 488 and 466 nm. The spectra were normalized by the main mode $E_{2H}$. (b) Room temperature resonant Raman scattering spectra of the samples excited by a 325 nm laser line. The luminescence background was subtracted and spectra were normalized by the first overtone (1LO). The inset shows the Raman spectra before the subtraction of the luminescence superimposed emission.

The crystallinity and the order of the induced structural defects in our samples can be further evaluated by the electron-phonon interaction, which is very sensitive to the atomic scale disorder. The electron-phonon coupling constant (α) can be probed under resonant excitation by both the number of the observed overtones (*n*) [34] and the ratio between the second and the first overtone ($I_{2LO}/I_{1LO}$) [35]. Post annealed ZnO-based materials with consequent improvement of the crystal quality have shown an increase of *n* and the $I_{2LO}/I_{1LO}$ ratio [36, 37]. Figure 3(b) presents the obtained spectra for our set of samples after subtraction of a broad luminescence peaking around 361 nm (3.43 eV) (Inset of Fig. 3(b)). Under resonant Raman condition the LO phonon overtones dominate due to their polar symmetry via intrinsic Frölich interaction with the excited electrons. A double Lorentzian fit of each overtone in the acquired spectra reveals central frequencies corresponding to multiple values of 572 cm$^{-1}$ for the $A_1$(LO) mode and 579 cm$^{-1}$ for the $E_1$(LO) mode. We notice in Fig. 3(b) a relatively high number of overtones, clearly *n* = 8 for the sample *AP*, and $n = 7$ for the heat treated samples. In addition, the ratio $I_{2LO}/I_{1LO}$, presented in Table 2, is larger than unity for the whole set of samples. Such large values for $n$ and $I_{2LO}/I_{1LO}$ indicate a high electron-phonon coupling strength and reveal the high crystallinity (high quality) of the studied samples. We also call attention to the fact that $n$ and $I_{2LO}/I_{1LO}$ gradually decrease with the heat treatment, corresponding to a reduction of electron-phonon coupling due to the introduction of lattice defects into the *w*-ZnO lattice.

**Table 2.** Wavenumber $\omega$ and $\lambda$ (in cm$^{-1}$) of the *w*-ZnO main mode $E_{2H}$, relative intensities of LO ($I_{LO}/I_{E2H}$) and S' ($I_{S'}/I_{E2H}$) to $E_{2H}$ mode, and the ratio $I_{2LO}/I_{1LO}$.

| Sample | $\omega$ | $\lambda$ | $I_{S'}/I_{E2H}$ | $I_{LO}/I_{E2H}$ | $I_{2LO}/I_{1LO}$ |
|---|---|---|---|---|---|
| *AP* | 438.01(6) | 8.4(3) | 2.7(5) | 1.4(2) | 2.55(7) |
| *H*600 | 436.72(6) | 9.8(2) | 3.1(4) | 2.1(2) | 2.05(3) |
| *H*800 | 436.33(5) | 10.9(3) | 3.8(7) | 2.9(3) | 1.82(3) |

### 3.4 X-ray Absorption Spectroscopy

Figure 4(a) shows the XANES spectra at Co *K*-edge obtained at room temperature for our polycrystalline $Zn_{0.96}Co_{0.04}O$ samples and for reference Co oxides, CoO (oxidation state +2) and $Co_2O_3$ (+3), and metallic Co (0). The comparison between the spectra from our samples and the references indicates that Co assumes predominantly the +2 oxidation state, corroborating the XRD results. We also observed the presence of a pre-edge peak in the spectra. As we have previously discussed [24, 38], this is only possible if the site where Co is located does not have an inversion center of symmetry, as in a tetrahedral configuration. In this sense, the observation of the pre-edge peak and the oxidation state +2 for the Co atoms in all samples are a strong indication that the $Co^{2+}$ ions are replacing the $Zn^{2+}$ in the *w*-ZnO structure.

We do not observe any significant changes in the pre-edge region due to the heat treatment, however we notice differences at the white-line peak. The upper-left inset of Fig. 4(a) highlights the differences between the recorded spectra at the white-line peak for the analyzed samples. With the heat treatment ($H$600) and its strengthening via increasing the temperature ($H$800), the white-line peak increases and shifts to higher energies. The lower-right inset of Fig. 4(a) shows the white-line peak of calculated XANES spectra for $Zn_{0.96}Co_{0.04}O$ samples using our XRD data obtained from Rietveld analysis and *ab initio* FEFF9 code [18]. The input files for FEFF9 code with cluster radius of 6.0 Å were generated using MAX-Crystalffrev software [16] varying the occupation rate of Zn and O sites. In this inset, the spectra labeled as $Zn_{0.50}$, $Zn_{0.75}$, $O_{0.50}$ and $O_{0.75}$ represent the calculated XANES spectra with occupation rate of 0.50 or 0.75 at Zn and O sites. The intensity and the energy of white-line peak increase as the occupation rate at Zn site decreases for the calculated spectra. On the other hand, the decrease of the occupation rate at the O sites results in the decrease of the white-line peak intensity and of its energy. Therefore, the calculated XANES spectra indicate that the observed increase in the intensity and the shift to higher energies of the white-line peak can be associated with the decrease of occupation rate at Zn site. In other words, we can conclude that the observed changes in the white-line peak are direct evidences of the introduction of vacancies at the Zn sites ($V_{Zn}$) due to the performed heat treatments.

Figure 4(b) shows the modulus of $k^3$ weighted Fourier transform (FT) extracted from Co $K$-edge for the $Zn_{0.96}Co_{0.04}O$ samples, a Co foil and Co oxides powders, as well as a spectrum obtained at Zn $K$-edge for the undoped ZnO reference sample. Qualitatively the spectra reveal that the crystallographic ambient of the $Co^{2+}$ ions in the set of samples is the same, and that the heat treatment does not promote any change on it. By comparison we also see that the spectra of the studied samples are quite different from those obtained for metallic Co, CoO and $Co_2O_3$ and, otherwise, they are quite similar to the spectrum acquired for the undoped ZnO reference sample at Zn $K$-edge. These observations led us again to conclude that the $Co^{2+}$ ions in all samples, despite the heat treatment, are located in the sites of the $Zn^{2+}$ ions in the *w*-ZnO lattice. We also performed a theoretical simulation of the measured FT for our samples. The technical details of the theoretical analysis were described in Ref. [39]. It was considered single and multi-scattering paths corresponding to the four successive atomic shells around Co substitutionally placed at the Zn sites of the ZnO structure according to the hexagonal wurtzite with P6$_3$mc space group. The parameters obtained via simulations, interatomic distances (*R*), coordination number (*N*), Debye-Waller factor ($\sigma^2$) and quality factor (*QF*) are shown in Table 3.

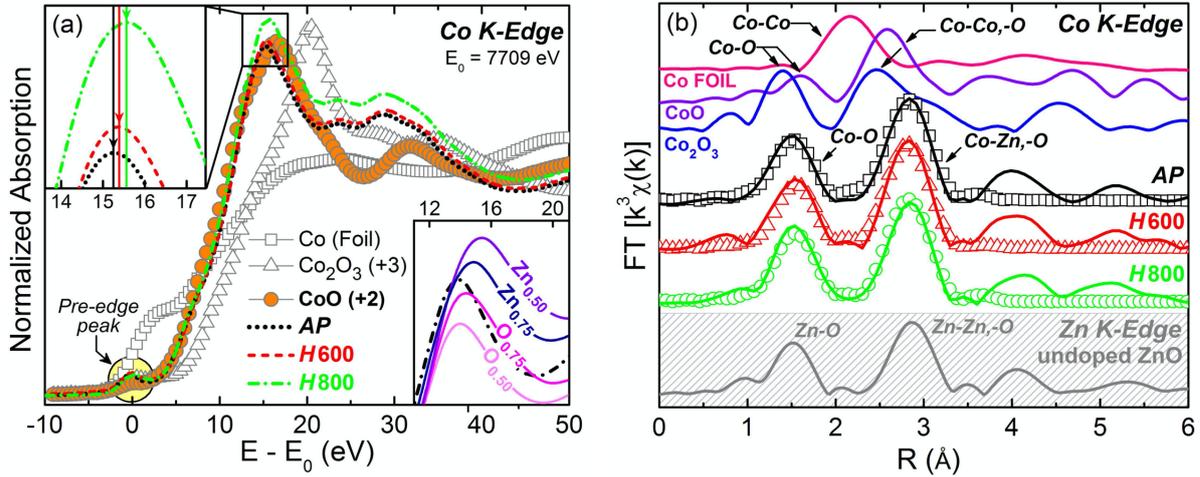

**Figure 4.** (a) Experimental Co *K*-edge XANES spectra for the whole set of samples (E$_0$ = 7709 eV). Spectra of metallic Co, rocksalt CoO (valence +2) and Co$_2$O$_3$ (valence +3) are also shown for comparison. The upper-left inset highlights the changes in the white-line peak. The lower-right inset presents the calculated white-line peak as a function of Zn and O occupation rate with respect to the full structure (black dashed-dot line). (b) $k^3$ weighted Fourier transforms of Co and Zn *K*-edge data of studied samples and references (Co, CoO and Co$_2$O$_3$). The spectra are offset for clarity. Here the lines correspond to the experimental data and the symbols correspond to the theoretical simulation results.

The obtained *QF* indicates the reliability of the fits, which is confirmed by the comparison of the fitted (symbols) and experimental (lines) spectra in Fig. 4(b). These data also confirm that the Co$^{2+}$ ions in the studied set of samples are taking place of the Zn$^{+2}$ ions in the *w*-ZnO structure (substitutional doping). An important and additional information can be extracted from the coordination number (*N*). For the Co-Zn shells *N* decreases substantially from sample *AP* to sample *H*600 and, finally, to the sample *H*800. This result, associated with the theoretical analyses of the changes in the white-line peak in the XANES spectra, indicates an increase of the Zn vacancies as a function of the heat treatment. On the other hand, no alterations, within the uncertainty of the calculation, are observed in the average coordination number for Co-O shells, which would be related to defects at the O sites [40].

**Table 3.** Co *K*-edge EXAFS simulation results obtained by assuming Co substitutionally placed at Zn sites in the *w*-ZnO matrix. *R*, with uncertainty 0.01 Å, is the distance from the central atom, *N* is the average coordination number, $\sigma^2$ the Debye-Waller factor and *QF* the quality factor [41].

| *Sample* | *Shell* | *R* (Å) | *N* | $\sigma^2$ (×10⁻³ Å²) | *QF* |
|---|---|---|---|---|---|
| *AP* | Co-O | 1.96 | 4.8(5) | 5(1) | 2.36 |
| | Co-Zn | 3.23 | 10.0(7) | 10.4(4) | |
| | Co-Zn | 3.20 | 5(2) | 10.4(4) | |
| | Co-O | 3.77 | 12(1) | 5(1) | |
| *H*600 | Co-O | 1.97 | 4.5(5) | 4(1) | 2.01 |
| | Co-Zn | 3.21 | 9(1) | 3.6(3) | |
| | Co-Zn | 3.34 | 6.1(3) | 3.6(3) | |
| | Co-O | 3.82 | 9(1) | 4(1) | |
| *H*800 | Co-O | 1.97 | 4.6(4) | 5(1) | 3.23 |
| | Co-Zn | 3.20 | 6.1(9) | 3(1) | |
| | Co-Zn | 3.33 | 6.0(7) | 3(1) | |
| | Co-O | 3.81 | 9.7(8) | 4(1) | |

### 3.5 *Photoluminescence and Electrical Characterization*

The structural defect identification and the evaluation of their relative densities were carried out through photoluminescence (PL) measurements. Figure 5(a) presents the low temperature (5 K) PL spectra obtained for the samples. The data were normalized by the near band edge (NBE) emission peak around 3.366 eV. We notice that with the heat treatment a low energy broadband evolves due to the introduction of defect levels inside the band gap, confirming the findings from the Raman measurements. The intensity of the defect band emission has a direct correlation with the strength of the heat treatment process and can be used as an indirect measurement of the relative density of defects introduced into the samples. For the *AP* sample the defect broad band cannot be seen, also pointing the high quality of the prepared samples. With the heat treatment (*H*600) a small defect broad band appears, and with the strengthening of the heat treatment (*H*800) the defect broad band starts to dominate the PL spectrum of the sample.

The inset in Fig. 5(a) shows the NBE region for the *AP* and the reference undoped ZnO samples (in logarithm vertical scale to highlight the low intensity features of the spectra). We identify three main emissions at the NBE region. The most intense emission at 3.366 eV is associated with excitons bound to neutral donors defect states ($D^0X$) recombinations, the relative small right shoulder around 3.37 eV is associated with free excitons (X) [42]. Only for the doped samples, besides the $D^0X$ and X emissions, we observe the presence of an additional line at 3.311 eV ($eA^0$). This emission is often observed and reported in the literature and is addressed to a transition of electrons from the conduction band to holes localized at relatively shallow-acceptor states related to specific structural defects [43]. Since it is not observed for the undoped sample we can infer that the related structural defects in question are those introduced in the *w*-ZnO lattice due to the Co-doping. Following these two emissions ($D^0X$ and $eA^0$) we also observed a sequence of longitudinal optical (LO) phonon replicas, corresponding to approximately 70 meV below the energy of the emissions [42].

The Gaussian fit of the defect broad band led us to associate it to two main different defect emissions, a band centralized at 2.501 eV (green luminescence) and another at 3.051 eV (violet-blue luminescence), Fig. 5(b) and (c). It is important to notice the relative intensities between these emissions. With the heat treatment the green emission starts higher in intensity than the violet-blue one (Fig. 5(b)), but as the heat treatment is intensified the violet-blue emission becomes dominant (Fig. 5(c)). The indexation of a band emission to a specific defect is a quite hard and controversial issue. For a long time, the green luminescence was indexed to oxygen vacancies ($V_O$), however a very recent experimental report has undoubtedly attributed this emission to zinc vacancies ($V_{Zn}$) [44, 45]. This indexation is also supported here by the experimental results and theoretical analysis obtained via X-ray absorption spectroscopy, there is no evidence of the presence of $V_O$. By its turn, following the representative theoretical [46, 47] and experimental reports [48–49], we index the violet-blue emission to interstitial zinc defects ($Zn_i$). Despite the difficulties to address the character of the defects in large band gap semiconductors, there is a consensus that $V_{Zn}$ is a shallow-acceptor type of defect, whereas $Zn_i$ is a shallow-donor type [45]. The character of these defects (acceptor or donor) was further confirmed by our first-principles DFT calculations.

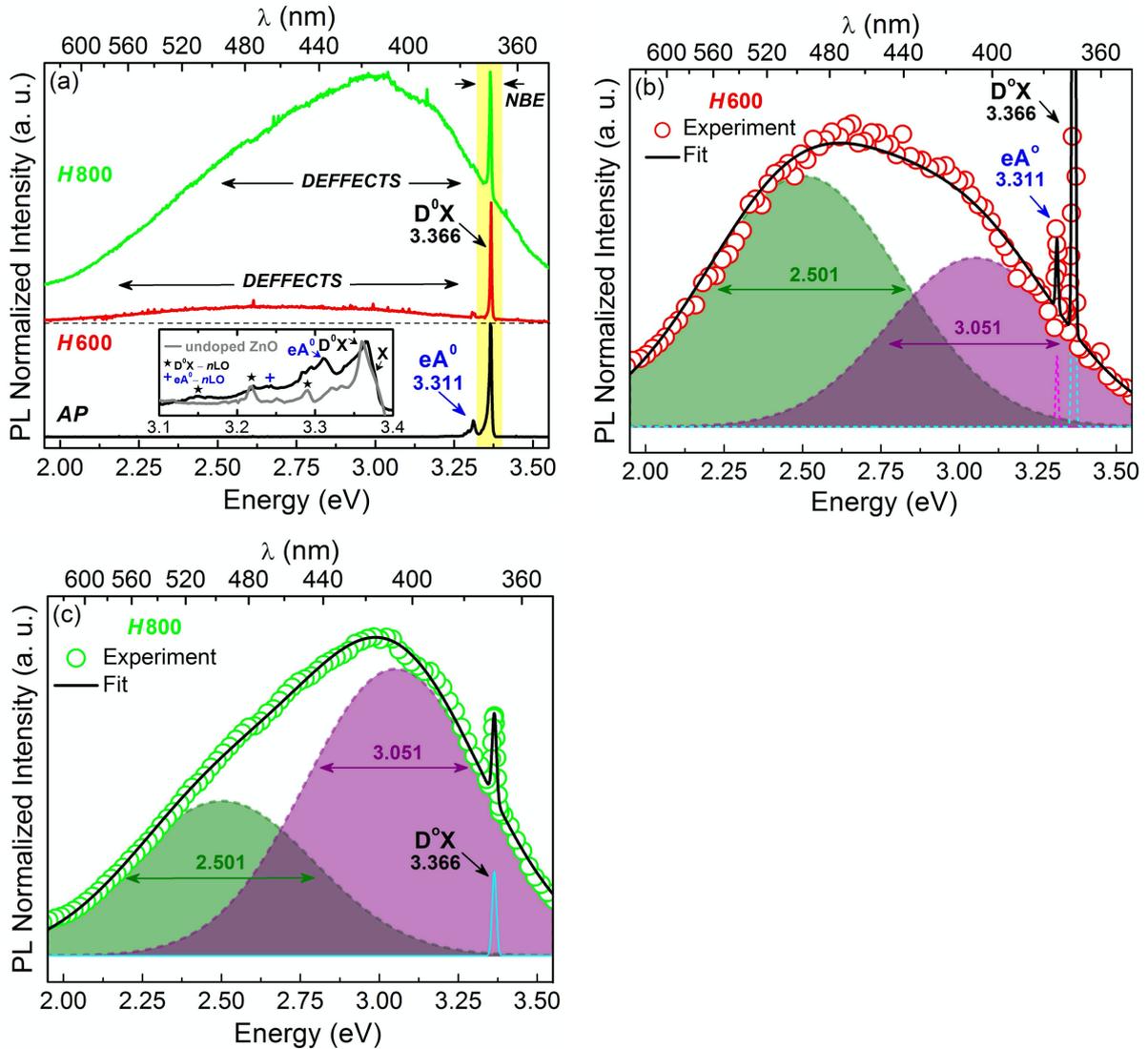

**Figure 5.** (a) PL spectra of the polycrystalline $Zn_{0.96}Co_{0.04}O$ samples. Measurements were performed at 5 K and excited with the wavelength of 325 nm of a He-Cd laser. The spectra are normalized by the integrated area of the NBE emission for comparison and offset for clarity. (b) and (c) present the experimental spectra and the Gaussian fits for the heat treated samples, *H*600 and *H*800, respectively.

To corroborate the PL results and the defect indexation we have also performed a *dc* electrical characterization. The results obtained via Hall effect measurements for the undoped ZnO reference sample and for our set of samples are presented in Table 4. All the samples are *n*-type semiconductors, but the most important results are the changes in the electron densities (*n*). As compared to the undoped ZnO reference sample, the doped ones present lower electron densities.

For the set of heat treated samples, the electron density of *H*600 is lower than that of *AP*, and for *H*800 the electron density assumes a value higher than those obtained for both *AP* and *H*600. The decrease of electron density with Co-doping is a well known fact, Co- and Mn-doping in ZnO introduce compensating defects to the system leading to the decrease of the electron density [50, 51]. The observation of the eA$^0$ emission related to shallow-acceptor defects for the doped samples is also an indication of these trends. The changes in the electron density of the heat treated samples can be understood in terms of the acceptor/donor character of the introduced defects $V_{Zn}$/Zn$_i$. As it can be seen from the PL measurements, for sample *H*600 there are more $V_{Zn}$ (acceptor defect) introduced into the system than Zn$_i$ (donor defect), as a consequence, we observe a compensation of the intrinsic donors leading to the decrease of electron density. On the contrary, for sample *H*800, by increasing the temperature of the heat treatment, more Zn$_i$ (donor defect) than $V_{Zn}$ (acceptor defect) is introduced into the system, which leads now to an increase of the electron density. Therefore, the changes in the electron density give an indication of the correct indexation of the observed defect band emissions in the PL measurements.

The measured Hall mobility ($\mu_H$) supports the above observed changes in the carrier density. From the reference undoped ZnO sample to the *AP* sample the mobility increases one order of magnitude, with the heat treatment (*H*600) the mobility increases more one order of magnitude, presenting a small reduction with the strengthening of the heat treatment (*H*800). The electron Hall mobility in polycrystalline non-degenerated semiconductors, bulk and thin films, are dominated by grain boundary scattering [52]. In this regime, for large grains and relative low carrier concentrations, the mobility is given by [53, 54]: $\mu=\mu_0\exp[-E_b/k_BT]$. Here $\mu_0=eL/(2\pi m^* k_B T)^{1/2}$ is the mobility inside the grain; where $L$ is the grain size, $m^*$ is the effective mass in the conduction band, $T$ the sample temperature, and $e$ and $k_B$ are the elementary charge and the Boltzmann constant, respectively. $E_b=e^2L^2n/8\epsilon$ is the energetic barrier height at the grain boundary; here $\epsilon$ is the dielectric permittivity, and $n$ is the carrier density in the bulk of the grain. Sintered cobalt doped ZnO samples present higher grain size values as compared to undoped ZnO samples. This fact is not yet well understood; however, in our case, a good explanation can be given by considering a liquid-phase sintering model. Sintering of pure ZnO is controlled by $Zn^{2+}$ solid-state diffusion [55]. Adding $Co_3O_4$, the sintering of ZnO at 1400 °C becomes liquid-phase assisted, since $Co_3O_4$ melts at 895 °C. The liquid enhances the $Zn^{2+}$ diffusion process and the grains grow rapidly according to a solution-precipitation phase-boundary reaction mechanism [55]. Therefore, the increase in the grains size and the decrease in the carrier concentration for the doped samples lead to an effective increase in the Hall mobility as compared to the undoped ZnO reference sample. With the heat treatment the grain size of the doped samples does not change substantially, and the Hall mobility becomes dominated by the carrier density. For sample *H*600 the carrier density (*n*) decreases, leading to an increase of the Hall mobility. With the strengthening of the heat treatment (sample *H*800), the carrier density increases, and, in spite of a possible improvement of the sintering process (small enlargement of the grains), the mobility now decreases.

**Table 4.** Hall mobility ($\mu_H$) and carrier density (*n*) at room temperature. Parameters obtained from the analysis of the *dc* electrical conductivity as a function of the temperature: activation energy ($E_A$), and the Mott parameters for VRH, the localized DOS at Fermi level ($N(E_F)$), the inverse localization length between localized states ($\delta$), parameter $\delta R$ after the average hopping distance (*R*), and average hopping energy (W). *R* and *W* were evaluated at 10 K ($k_B T = 0.86$ meV)

| Sample | $\mu_H$ (cm²/Vs) | $n$ ($10^{17}$cm$^{-3}$) | $E_A$ (meV) | $N(E_F)$ (cm$^{-3}$ eV$^{-1}$) | $\delta$ (cm$^{-1}$) | $\delta R$ | $W$ (meV) |
|---|---|---|---|---|---|---|---|
| undoped ZnO | 0.1 | 17.5 | 44.0 | $2.81 \times 10^4$ | 8.25 | 1.70 | 0.98 |
| *AP* | 2.3 | 1.3 | 31.1 | $9.50 \times 10^7$ | 220 | 2.61 | 1.50 |
| *H*600 | 49 | 0.8 | 28.3 | $2.80 \times 10^9$ | 353 | 1.60 | 0.92 |
| *H*800 | 44 | 4.4 | 16.6 | $6.49 \times 10^{14}$ | 24600 | 1.75 | 1.01 |

The formation of a defect band can be envisaged also via electrical transport studies. Temperature dependence of the *dc* electrical conductivity ($\sigma$) of all studied samples presents a typical semiconductor behavior, as the temperature decreases, the conductivity decreases exponentially. Figure 6 shows the Arrhenius plot (ln$\sigma$ *vs.* 1/*T*) for all samples. It is observed that the experimental data at higher temperatures can be readily linearly fit with no exceptions. This behavior indicates that the conduction mechanism at higher temperatures is governed by thermally activated type of band conduction. The activation energy ($E_A$) obtained from the linear fit is also presented in Table 4. We observe a systematic decrease of $E_A$ with the Co-doping and the heat treatment. This is interpreted as an evolution of donor/acceptor defect band close to the conduction/valence band with Co-doping and the subsequent heat treatment, resulting in the observed decrease in $E_A$. In the range of low temperatures ($T$ <50 K) most of the free electrons/holes in the conduction/valence band become trapped in the donor/acceptor defects levels below/above the conduction/valence band, which means that delocalized carriers become localized. In this situation carriers hop from an occupied localized state to an unoccupied localized one within the defect band. It is the variable range hopping (VRH) model proposed by Mott [56]. In the Mott's VRH model, a relationship between the conductivity σ and the temperature $T$ is:

$$\sigma = \sigma_0 \exp[-(T_0/T)^{1/4}], \qquad (1)$$

here $\sigma_0$ and $T_0$ are expressed functionally as [57]:

$$\sigma_0 = 3e^2 \nu_{ph}/(8\pi k_B)^{1/2} \times [N(E_F)/\delta T]^{1/2}, \qquad (2)$$

$$T_0 = 16\delta^3/k_B N(E_F), \qquad (3)$$

where $\nu_{ph}$ is the phonon frequency at Debye temperature ($\approx 10^{13}$ Hz), $N(E_F)$ is the density of localized states for electrons at the Fermi level and δ the inverse localization length of a wave function for localized state. From equations 2 and 3 we deduce that $\ln(\sigma T^{1/2}) \propto T^{-1/4}$. The insets in Fig. 6 show the plot of $\ln(\sigma T^{1/2})$ *vs.* $T^{-1/4}$, the linear behavior at temperatures below 50 K indicates a dominant VRH type of conduction is this temperature range. To further evaluate this condition one also has to consider two other parameters: the hopping distance $R=[9/(8\pi\delta k_B T N(E_F))]^{1/4}$, and the average hopping energy $W=3/[4\pi N(E_F) R^3]$. For VRH type of conduction the value of δ$R$ is required to be greater than 1, whereas $W$ must be greater than $k_B T$. The extracted parameters δ, $N(E_F)$, δ$R$ and $W$ are presented in Table 4. First, we note that the requirements δ$R$ > 1 and $W > k_B T$ are completely fulfilled for all the samples. We call attention to the obtained relative low values of the inverse localization length ($\delta$) and the density of localized states at the Fermi level ($N(E_F)$) for the undoped ZnO reference sample as compared to the values obtained for our Co-doped ZnO. With the Co-doping and the heat treatment the localization ($1/\delta$) and the $N(E_F)$ increase, confirming the emergence of defect bands close to the edges of the valence and conduction bands. Specially, the values for $N(E_F)$ and $\delta$ for sample *H*800 are 5 and 2 orders of magnitude higher than those for sample *H*600, respectively. And the most important, these differences are in straight correlation with the differences in the relative intensities observed for the PL defect broad band emission presented before (Fig. 5(a)). Nevertheless, the values obtained for $N(E_F)$ and $\delta$ for our set of samples are also quite low as compared with the results obtained for the TM-doped ZnO in other reports [58, 59], revealing also the high quality of our samples.

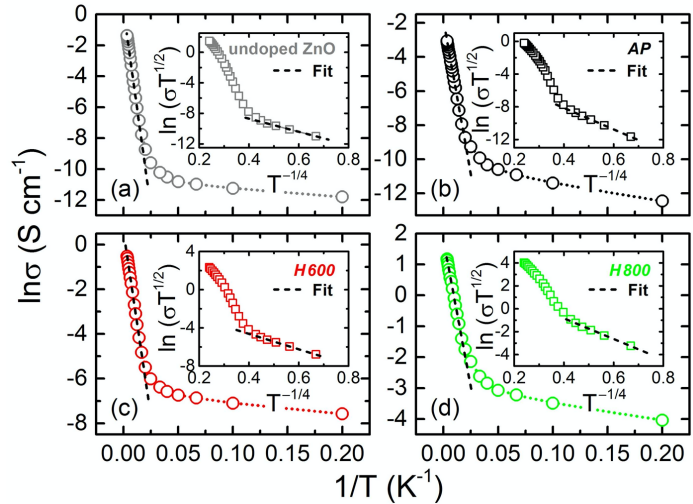

**Figure 6.** Arrhenius plot of *dc* conductivity (ln$\sigma$ *vs.* $1/T$) for the (a) reference undoped ZnO, (b) *AP*, (c) *H*600 and (d) *H*800 samples. The insets present the plot $\ln(\sigma T^{1/2})$ *vs.* $1/T^{-1/4}$. The symbols are the experimental data and the dashed lines correspond to linear fit to the data in the correspondent temperature range.

3.6 *Magnetic Characterization*

The magnetic measurements were performed in a Cryogenics superconducting quantum interference device (SQUID) system under magnetic fields up to 60 kOe in the range of 50 to 300 K. Figure 7(a) presents the obtained first magnetization curves (*M-H*) at room temperature. The result for the *AP* sample reveals a pure paramagnetic response, as the same reported previously [12]. For the *H*600 sample we observe the emergence of a small ferromagnetic phase superimposed to the paramagnetic phase. Increasing the temperature of the heat treatment for the *H*800 sample the ferromagnetic phase is now clearly seen dominating the magnetic behavior besides the paramagnetic phase common to the three studied samples. Figure 7(b) shows the *s*-shaped ferromagnetic hysteresis after subtraction of the paramagnetic phase.

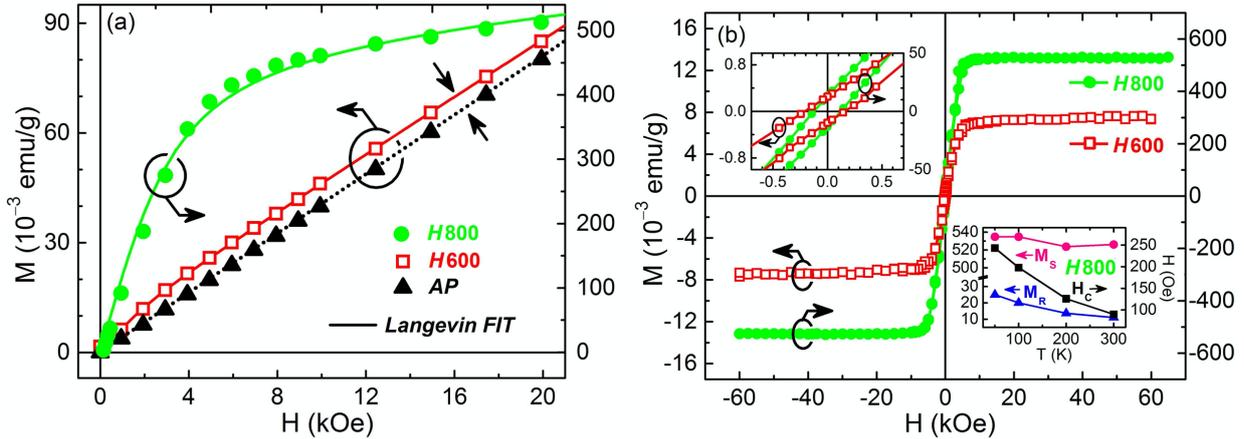

**Figure 7.** Results of the magnetic characterization performed at 300 K. (a) *M-H* first magnetization curves for the studied set of samples. The solid lines correspond to the fit with Langevin function plus a term corresponding to the paramagnetic response (Eq. 4). (b) Hysteresis loop for the samples *H*600 and *H*800 after subtracting the paramagnetic component. The insets show the coercive field around 0.01 T (100 Oe) for the samples *H*600 and *H*800 (upper-left) and the magnetic parameters (saturation magnetization ($M_S$), remanent magnetization ($M_R$), and coercive field ($H_C$)) as a function of the temperature (lower-right).

Here it is important to stress that the magnetic results are in straight correspondence to the density of structural defects ($Zn_i$ and $V_{Zn}$) introduced into the samples via the heat treatment, leading us to interpret these results under the scope of the static magnetic polaron theory of Coey *et al.* [13]. In this model, the hybridization between localized carriers trapped in a defect, a polaron, and the dopant atoms leads to ferromagnetic exchange coupling between them, forming what is called a bound magnetic polaron (BMP). In fact, cobalt atoms can hybridize with shallow-donor defect bands in *w*-ZnO [60]. In the literature one can find a large number of reports on the observed ferromagnetism for doped *w*-ZnO explained in terms of the BMP theory by considering oxygen vacancies ($V_O$) as the necessary shallow-donor defect associated with the formation of the polarons. However, as we have already pointed out [61], $V_O$ cannot be considered due to its deep-donor character [46, 62]. On the other hand, $Zn_i$ is a better candidate since we only observed the presence of $V_{Zn}$ (acceptor defect) and $Zn_i$ (donor defect) introduced in the systems via the heat treatment, and no signs of $V_O$ are experimentally detected. This assumption is also corroborated by our DFT results presented ahead.

To quantify the magnetic parameters of the two coexisting phases, we followed the magnetic analysis described in Refs. [61, 63]. The obtained magnetic parameters are listed in Table 5. First, we call attention to the paramagnetic susceptibility ($\chi_P$). From sample *AP* to the heat treated sample *H*600, and with the increasing of the temperature of the heat treatment for sample *H*800, the $\chi_P$ decreases, corresponding to a growing number of ferromagnetically coupled Co atoms. $N_{BMP} = \mu/M_S$ is the density of ferromagnetic units, the density of BMPs. We observe that $N_{BMP}$ increases two orders of magnitude from sample *H*600 to *H*800. As pointed out by Coey *et al.*, the percolation threshold of BMPs for the appearance of long-range ferromagnetic order in ZnO is of the order of $10^{19}$/g [13]. Therefore, for the sample *H*600 the $N_{BMP}$ is below the threshold limit, which leads to a partial fulfilling of the volume of the sample with BMPs, leaving a fraction of the Co atoms uncoupled ferromagnetically (higher $\chi_P$). With the increase of the temperature of the heat treatment (*H*800), $N_{BMP}$ reaches the threshold limit, more Co atoms are now ferromagnetically coupled (lower $\chi_P$). However, the observation of a paramagnetic component even for the *H*800 sample means that the calculated BMP percolation threshold by Coey *et al.* [13] is underestimated for our samples. Table 5 also presents the magnetic moment ($\mu$) per BMP; the obtained values are much larger than the reported values for any Co related structured system: Co metal (1.72 $\mu_B$/Co), nanostructured Co cluster (2.1 $\mu_B$/Co) [64, 65]. High moments per cation, like the ones reported here, cannot be explained in terms of possible known ferromagnetic phases [66, 67]. Thus, it is an evidence that the observed ferromagnetism is not a function of secondary ferromagnetic phases. Besides, it can be accounted to the average sum over the individual magnetic moments of the Co atoms within the BMP.

The magnetic data can be also analyzed via the BMP model as described by McCabe *et al.* [68]. According to this model, the measured magnetization can be fitted at relatively higher temperatures to the relation:

$$M = N_{BMP} m_S L(m_S H/k_B T) + \chi_P H. \tag{4}$$

Here the first term is the BMP contribution and the second term is the paramagnetic contribution. $N_{BMP}$ is the density of BMPs, $m_S$ is the spontaneous magnetic moment per BMP. $L(x)=\coth(x)-1/x$ is the Langevin function. The Langevin fits to the first magnetization curves are also presented in the Fig. 7(a). The parameters $N_{BMP}$ and $m_S$ obtained from the fit are $1.91 \times 10^{17}$/g and 4.57 $\mu_B$, respectively, for sample $H$600 and $1.67 \times 10^{19}$/g and 3.22 $\mu_B$ for sample $H$800. These values are in quite good agreement with the previous analysis confirming the assumption of the ferromagnetic order observed in our samples explained in terms of the formation of BMPs.

**Table 5.** Parameters of the magnetic characterization: saturation magnetization ($M_S$), paramagnetic susceptibility ($\chi_P$), density of ferromagnetic units ($N_{BMP}$) and estimated magnetic moment per BMP ($\mu$).

| Sample | $\chi_P$ ($\times 10^{-6}$ emu/gOe) | $M_S$ (emu/g) | $N_{BMP}$ (g$^{-1}$) | $\mu$ ($\mu_B$) |
|---|---|---|---|---|
| $AP$ | 4.03 | - | - | - |
| $H$600 | 3.89 | 0.0073 | $1.82 \times 10^{17}$ | 4.33 |
| $H$800 | 2.82 | 0.5150 | $1.60 \times 10^{19}$ | 3.53 |

### 3.7 First-Principles Calculations

Based on the experimental results described in the previous sections, first-principles calculations were carried out in order to study the origin of magnetism and the role of intrinsic defects in the magnetic properties of the Co-doped $w$-ZnO samples. We considered Zn interstitial ($Zn_i$) and Zn vacancies ($V_{Zn}$) defects, as well as Co doping at Zn sites and the interaction of the Co impurities with the intrinsic defects in the $w$-ZnO lattice, i.e., Co with $Zn_i$, and Co with $V_{Zn}$. Figure 8 shows the total calculated spin-resolved electronic density of states (DOS) for the systems investigated here: undoped pristine $w$-ZnO (Fig. 8(a)), undoped $w$-ZnO with $Zn_i$ defect (Fig. 8(b)), undoped $w$-ZnO with $V_{Zn}$ (Fig. 8(c)), Co-doped $w$-ZnO (substitutional Co dopant at the Zn site) without defect (Fig. 8(d)), Co-doped $w$-ZnO with $Zn_i$ (Fig. 8(e)), and Co-doped $w$-ZnO with $V_{Zn}$ (Fig. 8(f)). For the undoped pristine $w$-ZnO (Fig. 8(a)) we observe a typical semiconductor DOS with no spin-polarization. With the introduction of the $Zn_i$ defects, the valence band is relatively pulled down and the conduction band is stretched until crossing the Fermi level (Fig. 8(b)), but no spin-polarization is observed for this case. On the contrary, with the $V_{Zn}$ defect (Fig. 8(c)) the valence band is pulled up and spin-polarization is observed mainly near the Fermi level leading to a net magnetic moment of 1.55 $\mu_B$/cell. We can infer here that $V_{Zn}$ defects may account for the often observed ferromagnetism in undoped w-ZnO [24]. The crossing of the Fermi level by the conduction band in the undoped w-ZnO with $Zn_i$ and by the valence band in the undoped w-ZnO with $V_{Zn}$ indicates that $Zn_i$ and $V_{Zn}$ have donor and acceptor characters, respectively, confirming the previous assumptions about the indexation of the defects band emissions in the PL results and the carrier density changes for the studied samples under heat treatment. For the Co-doped w-ZnO system without defect (Fig. 8(d)), we observe the emergence of an impurity band inside the band gap with a large exchange spin-splitting, a fully occupied spin-up band and a partially occupied spin-down band, resulting, as expected, in a net magnetic moment of 3.37 $\mu_B$/cell. Considering now the interaction of the Co impurity with intrinsic defects in w-ZnO lattice, the obtained total DOS for Co-doped w-ZnO with $Zn_i$ defect is spin-polarized with a total magnetic moment of about 2.05 $\mu_B$/cell (Fig. 8(e)). Finally, the DOS of the Co-doped w-ZnO system with $V_{Zn}$ presents a large spin-polarization with a net magnetic moment of 4.93 $\mu_B$/cell (Fig. 8(h)), which is larger than the obtained net magnetic moment of w-ZnO systems with only $V_{Zn}$ defects (1.55 $\mu_B$/cell) or only Co impurities (3.37 $\mu_B$/cell). This is an indication of a ferromagnetic (FM) coupling between the magnetic moments due to $V_{Zn}$ and Co.

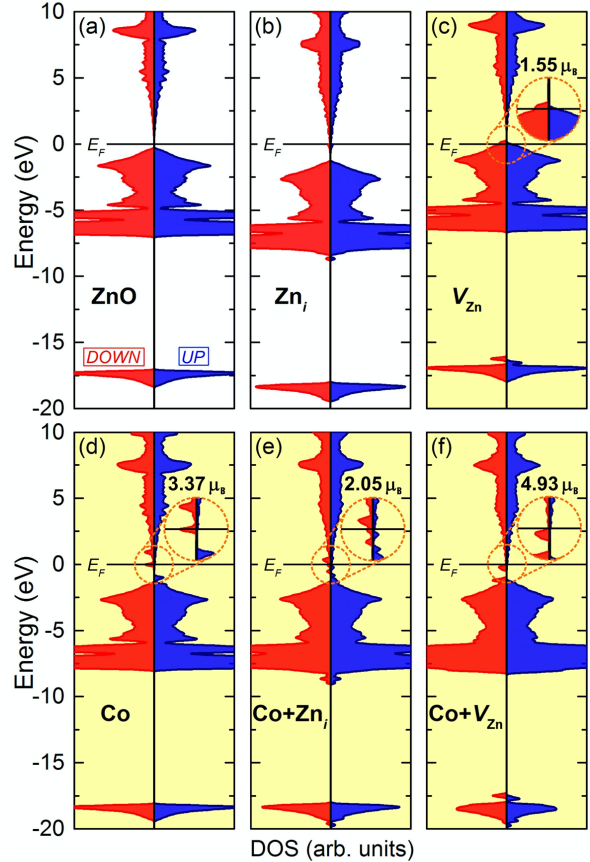

**Figure 8.** Total spin-resolved DOS for (a) undoped (pristine) *w*-ZnO, (b) zinc interstitial ($Zn_i$), (c) zinc vacancy ($V_{Zn}$), (d) for *w*-ZnO with a substitutional Co dopant at the Zn site (Co), and for *w*-ZnO with (e) Co and $Zn_i$ and (f) Co and $V_{Zn}$. Here the Fermi level is set to energy zero ($E_F = 0$ eV). The yellow panels correspond to those ones with nonzero net magnetic moment.

To gain more insight about the magnetic interaction between the Co impurities and the $Zn_i$ and $V_{Zn}$ defects, we calculated the electronic charge and the spin densities contours for each system. Figure 9(a) and (b) present the electronic charge density at a cross section of the Co-doped *w*-ZnO lattice containing the Co impurity and the $Zn_i$ and $V_{Zn}$ defects, respectively. We can see in Fig. 9(a) a distortion in the structure around the $Zn_i$. We found that the optimized $Zn_i$-O bond lengths are 1.91 Å and 1.88 Å, larger than the other Zn-O bond lengths in the rest of the lattice. The optimized Co-O bond lengths are 1.87 Å and 1.80 Å, which are smaller than that for the Zn-O in the *w*-ZnO lattice, 1.98 Å; such reduction in the bond lengths indicates an increase in the covalent character of the bonds in the *w*-ZnO. In *w*-ZnO with $V_{Zn}$ (Fig. 9(b)), we found that $V_{Zn}$ defect did not introduce any significant structural distortions into the *w*-ZnO lattice.

Considering the spin spatial distribution, Fig. 9(c) and (d) show the correspondent spin density contours for the *w*-ZnO lattice presented in Fig. 9(a) and (b), respectively. First, we call attention to that the magnetic moment is mainly located around the Co atoms in all cases. For the Co-doped *w*-ZnO lattice with $Zn_i$ (Fig. 9(c)) we observe that the positive spin-polarization around the Co atom spreads over the surrounding O atoms, which can be seen as a FM coupling between the Co and the O atoms, and can be explained by the increase in the covalent character of the bonds. The obtained local magnetic moments are 1.83 $\mu_B$ and 0.12 $\mu_B$ around the Co and O atoms, respectively. Besides, a small negative spin-polarization around the $Zn_i$ (-0.07 $\mu_B$) and over its left neighbour (-0.03 $\mu_B$) is also observed, which also indicates a FM coupling between the $Zn_i$ and the O atoms. Since the undoped *w*-ZnO system with $Zn_i$ does not present any spin-polarization (Fig. 8(b)), we conclude that this spin-polarization around the $Zn_i$ is induced by the Co atoms, and the coupling between the $Zn_i$ and the Co is AF, as can be clearly seen in Fig. 9(c). We also would like to call attention to other two points: first, no spin-polarization is induced on the Zn atoms located at the regular lattice in ZnO; second, the spin density is quite localized over the Co and the $Zn_i$, not spreading over the whole structure. For the Co-doped *w*-ZnO lattice with $V_{Zn}$ (Fig. 9(d)) we observe a positive spin-polarization mainly around the Co and the oxygen atoms surrounding the $V_{Zn}$, corresponding, as mentioned before, to a FM coupling between Co magnetic moment and the moment due to the $V_{Zn}$ in its surrounding oxygen atoms. We also observe a broader spreading of magnetic moments, even at the Zn atoms in the *w*-ZnO lattice. Here the obtained local magnetic moments at the Co impurity and at its the closest O atom are 2.82 $\mu_B$ and 0.24 $\mu_B$, respectively.

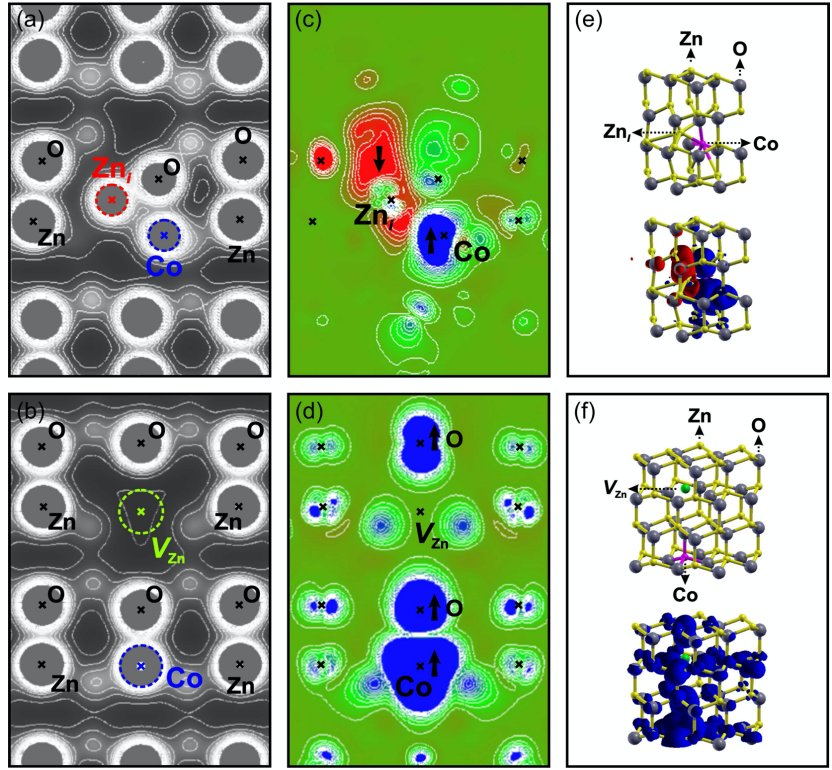

**Figure 9.** Calculated electronic charge density of a *w*-ZnO lattice cross section for Co-doped *w*-ZnO systems with (a) $Zn_i$ and (b) $V_{Zn}$ defects. Spin density contours for Co-doped *w*-ZnO systems with (c) $Zn_i$ and (d) $V_{Zn}$. (e) and (f) show the correspondent three-dimensional view of the structure and the spin spatial distribution in (a) and (c), and in (b) and (d), respectively.

Finally, we analyze the interaction between Co atoms into the *w*-ZnO lattice. Dealing with high quality (low level of defects) Co-doped *w*-ZnO samples we have already shown experimentally that the magnetic interaction between the Co atoms is AF [12, 39], exceptionally with the concomitant introduction of structural defects into the *w*-ZnO lattice [61, 63]. In previous theoretical calculations [24] we have also obtained AF coupling between two Co atoms in the *w*-ZnO lattice without defects, demonstrating the important role performed by structural defects in order to achieve the desired RTFM. In such a context, we carried out additional calculations with two Co atoms placed at the Zn sites in the *w*-ZnO lattice with $Zn_i$ and $V_{Zn}$ defects. The distance between the two Co atoms was varied from about 3 to 8 Å (limited by the size of the supercell, 16 Å) and their total energies in the FM and AF states in the condition of all atoms fully relaxed were calculated. We found that in the presence of $Zn_i$, for a separation distance of around 3 Å (high Co concentrations) the Co atoms interact antiferromagnetically. Whereas, the interaction turns to ferromagnetic for the larger separation (low Co concentrations), about 8 Å. This result can be interpreted as follows. As $Zn_i$ has a *n*-type character and introduces local structural distortions into the *w*-ZnO structure, we infer that it can form a polaron; in such a context the Co atom hybridizes with the shallow-donor defect states induced by $Zn_i$ leading to the formation of a magnetic polaron. Therefore, $Zn_i$ in *w*-ZnO lattice not only behaves as a *n*-type impurity, but also mediates the FM coupling between the Co atoms in systems with relative lower Co concentrations. It is worth stressing that our DFT results for Co-doped *w*-ZnO with $Zn_i$ can be fully understood under the BMP model for the experimentally observed ferromagnetic order [13]. In turn, when the same calculations are performed for $V_{Zn}$, we found that the two Co atoms only couple antiferromagnetically, independent of the Co-Co separation [24]. Therefore, in spite of the highest net magnetic moment achieved with the presence of $V_{Zn}$, 4.93 $\mu_B$/cell (Fig. 8(f)), Co-doped *w*-ZnO can be only a FM material with $Zn_i$.

## 4 CONCLUSION

In summary, we have presented a systematic structural, optical, electrical and magnetic characterization of Co-doped ZnO polycrystalline bulk samples ($Zn_{0.96}Co_{0.04}O$). The structural analysis confirms that Co atoms in the studied samples are occupying Zn-sites of the *w*-ZnO structure. Clearly, the results exclude the presence of magnetic extrinsic source such as Co-rich nanocrystals and segregated secondary magnetic phases. X-ray absorption analyses indicate the presence of $V_{Zn}$ in the samples, and increasing the temperature of the heat treatment, the $V_{Zn}$ concentration also increases. PL and electrical characterization reveal the formation of two defect bands close to the conduction and the valence band (shallowdonor and shallow-acceptor bands) associated with $Zn_i$ and $V_{Zn}$, respectively. The magnetic response observed for all the samples reveals a direct correlation with the introduction of the structural defects ($Zn_i$ and $V_{Zn}$) via heat treatment and the emergence of a room temperature ferromagnetic phase. These results lead us to evoke the BMP model to explain the obtained magnetic data with $Zn_i$ as the most promising defect of shallow-donor character necessary in the context of the BMP model to promote the desired ferromagnetic order in the ZnO:Co system. Finally, theoretical DFT calculations on the structural, electronic and magnetic properties of the Co-doped $w$-ZnO with $Zn_i$ and $V_{Zn}$ defects give a strong support for the assumptions asserted from the experimental data.

## ACKNOWLEDGMENTS

Support from agencies CNPq (grant 306715/2018-0) and FAPESP (07/56231-0; grant 2015/16191-5) is gratefully acknowledged. This research used resources of the Brazilian Synchrotron Light Laboratory (LNLS), an open national facility operated by the Brazilian Centre for Research in Energy and Materials (CNPEM) for the Brazilian Ministry for Science, Technology, Innovations and Communications (MCTIC). The XAFS2 beamline staff is acknowledged for the assistance during the experiments.